\documentclass[preprint,12pt]{elsarticle}
\usepackage[T1]{fontenc} 
\usepackage[letterpaper,top=2cm,bottom=2cm,left=3cm,right=3cm,marginparwidth=1.75cm]{geometry}
\usepackage{amsmath}
\usepackage{amssymb}
\usepackage{graphicx}
\usepackage{multirow}
\usepackage{booktabs}
\usepackage{longtable}
\usepackage{subcaption}

\usepackage[colorlinks=true, allcolors=blue]{hyperref}

\usepackage{amssymb}
\usepackage{amsmath}

\journal{...}

\begin{document}

\begin{frontmatter}



\title{Noise-Aware Optimization in Nominally Identical Manufacturing and Measuring Systems for High-Throughput Parallel Workflows}

\author[inst1]{Christina Schenk\corref{cor1}}
\ead{christina.schenk@imdea.org}
\author[inst1,inst2]{Miguel Hern\'andez-del-Valle}
\author[inst1,inst2]{Luis Calero-Lumbreras}
\author[inst3]{Marcus Noack}
\author[inst1]{Maciej Haranczyk\corref{cor1}}
\ead{maciej.haranczyk@imdea.org}
\affiliation[inst1]{organization={IMDEA Materials Institute},
	addressline={C Eric Kandel 2}, 
	city={Getafe},
	postcode={28906}, 
	state={Madrid},
	country={Spain}}
\affiliation[inst2]{organization={Universidad Carlos III de Madrid},
	            city={Leganés}, 
	            postcode={28911}, 
				state={Madrid},
				country={Spain}}
\affiliation[inst3]{organization={Applied Mathematics \& Computational Research Division, Lawrence Berkeley National Laboratory}, 
	city={Berkeley}, 
	state={CA},
	country={USA}}
\cortext[cor1]{Corresponding authors}

\begin{abstract}
Device-to-device variability in experimental noise critically impacts reproducibility, especially in automated, high-throughput systems like additive manufacturing farms. While manageable in small labs, such variability can escalate into serious risks at larger scales, such as architectural 3D printing, where noise may cause structural or economic failures. This contribution presents a noise-aware decision-making algorithm that quantifies and models device-specific noise profiles to manage variability adaptively. It uses distributional analysis and pairwise divergence metrics with clustering to choose between single-device and robust multi-device Bayesian optimization strategies. Unlike conventional methods that assume homogeneous devices or generic robustness, this framework explicitly leverages inter-device differences to enhance performance, reproducibility, and efficiency. An experimental case study involving three nominally identical 3D printers (same brand, model, and close serial numbers) demonstrates reduced redundancy, lower resource usage, and improved reliability. Overall, this framework establishes a paradigm for precision- and resource-aware optimization in scalable, automated experimental platforms.
\end{abstract}

%


\end{frontmatter}



\section{Introduction}
Recent advances in automation technologies have revolutionized scientific research, particularly in fields that rely on high-throughput experimentation. The ability to execute vast numbers of experiments in parallel with minimal human intervention has fundamentally accelerated the pace of scientific discovery, enabling the exploration of expansive parameter spaces critical to applications such as materials science, chemical synthesis, and sensor calibration \cite{lyall-brookes_flow_2025,wakabayashi_bayesian_2022,rankovic_bayesian_2024,GONZALEZ2023108110}. By embracing automation and parallelization, researchers achieve increased throughput, enhanced reproducibility, and more targeted higher data volumes, all of which are essential for rapid iteration and timely insights. However, as these automated systems scale from small laboratory setups to complex industrial or architectural applications, device-to-device variability in experimental noise can amplify significantly, posing a key barrier to precision and reliability that must be addressed to realize the full potential of automated, high-throughput workflows \cite{Holland2020, WIRTH2024104081}.

Yet, this drive to accelerate discovery brings fresh challenges to measurement reliability. Even when using nominally identical experimental platforms, subtle variations, stemming from instrumentation, environment, or fabrication, can induce markedly different noise characteristics across devices. This device-to-device variability can bias outcomes and undermine reproducibility \cite{li_measuring_2011,han_repeatability_2017,malyarenko_multi-system_2013, Holland2020,WEGENER2021611,HALL2025113588}. Addressing these issues by identifying, modeling, and mitigating noise in experimental systems is important \cite{sassella_noise_2022, Holland2020,WEGENER2021611}, particularly in the development of fully automated and high-throughput workflows. 

Early approaches to managing experimental noise have traditionally relied on repeated measurements and heuristic averaging, but this can become resource-prohibitive in high-throughput or fully automated settings \cite{sassella_noise_2022, slautin_measurements_2024}. More recent developments in Bayesian optimization and other probabilistic frameworks have enabled explicit modeling of noisy observations, improving the efficiency of parameter exploration under uncertainty \cite{wakabayashi_bayesian_2022,rankovic_bayesian_2024,letham2019constrainedbayesianoptimizationnoisy,slautin_measurements_2024}. Extensions of these frameworks have addressed batch \cite{bellamy2022} or parallel experiment design \cite{GONZALEZ2023108110}. Some methods now account for experiment failure or missing data through probabilistic imputation or explicit classification \cite{wakabayashi_bayesian_2022}. 

However, a majority of current strategies either assume all devices behave statistically identically or attempt to design robust solutions agnostic to device-specific variability, potentially under-utilizing opportunities for targeted optimization or risking bias if variability is significant \cite{li_online_2024}. Recent advances in distributed and online multi-task learning using the alternating direction method of multipliers (ADMM) offer promising paradigms to explicitly learn and exploit inter-device relationships in parallel optimization settings \cite{li_online_2024}. These approaches dynamically model correlations between tasks and are especially well-suited for decentralized experimental setups, where they overcome the limitations of centralized servers by enabling efficient local information exchange directly among experimental devices.

Our approach advances the state-of-the-art by explicitly characterizing and analyzing the distribution of noise across a set of seemingly identical parallelized measurement systems (i.e., same brand and model, close serial numbers). We introduce a noise-aware decision-making algorithm that, given the observed heterogeneity, determines whether to pursue single-device optimization or to adopt multi-device robust optimization. By adaptively tuning the optimization paradigm to the measured noise landscape, our method enhances efficiency when devices are indeed similar while maintaining robustness and reliability when variability is non-negligible. This principled, data-driven framework thus bridges the gap between device-specific optimization and agnostic robustness, unlocking higher throughput without sacrificing confidence in experimental outcomes.

\begin{figure}[h!]
	\centering
	\includegraphics[width=1\linewidth]{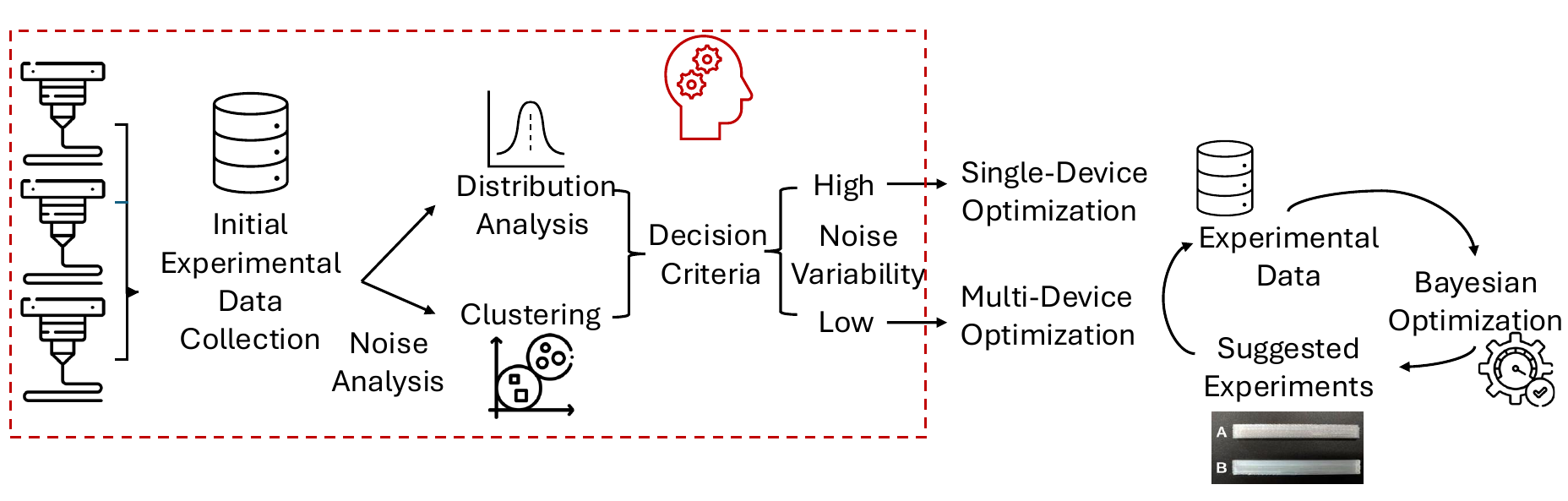}
	\caption{Workflow.}
	\label{fig:workflow}
\end{figure}

The proposed workflow, illustrated in Figure~\ref{fig:workflow}, begins with initial experimental data collection across multiple nominally identical devices, followed by comprehensive noise analysis and statistical distribution analysis of the observed measurement variability. Through clustering techniques, devices are grouped based on their noise characteristics, and the overall noise variability across the system is assessed. Based on predefined decision criteria, the algorithm determines whether the observed variability is sufficiently low to warrant multi-device robust optimization or high to enable single-device optimization. The chosen optimization strategy then guides Bayesian optimization to suggest the next set of experiments, creating an iterative feedback loop that continuously refines both the noise characterization and the experimental design. This adaptive framework ensures that the optimization strategy remains aligned with the actual noise landscape throughout the experimental campaign.

Integrating noise-awareness into optimization workflows ensures that increased \\throughput and automation do not come at the cost of accuracy and reproducibility, paving the way for more resilient and insightful automated discovery platforms.

The paper is organized as follows. Section~\ref{sec:meth} describes the methodologies used for experimental data acquisition and noise analysis. It introduces the optimization decision criterion designed to select between single-device and multi-device strategies, and concludes with a detailed explanation of the Bayesian optimization algorithms and implementation specifics. Section~\ref{sec:results} presents the initial experimental data collection, followed by noise analysis results, including distributional characterization, clustering, and pairwise divergence metrics. This section also reports on the settings of both single-device and multi-device Bayesian optimization approaches. In Section~\ref{sec:discuss}, these results are discussed in the broader context of high-throughput automated experimentation, focusing on scalability, reproducibility, and resource efficiency. The article concludes with a summary of the main findings in Section~\ref{sec:concl}.

\section{Methods}\label{sec:meth}
This section details the noise-aware decision-making algorithm, beginning with the experimental procedures and methods used to collect data and characterize noise variability, followed by the definition of the optimization decision criteria. It concludes with a description of the Bayesian optimization algorithms employed to evaluate single-device and multi-device strategies.
\subsection{Initial Experimental Data Collection}
To enable the core noise analysis underlying the presented noise-aware decision-making algorithm, it is essential to collect a sufficiently large initial dataset that thoroughly explores the parameter space. This dataset must include repeated experimental runs to robustly capture inherent variability and noise characteristics across devices, ensuring reliable statistical analysis and informed optimization decisions.

\subsection{Noise Analysis: Distribution Analysis, Clustering, and Pairwise Divergence Metrics}
To rigorously characterize the variability and noise inherent in the outputs of nominally identical manufacturing or measuring devices $n$, a comprehensive noise analysis is required using multiple complementary statistical techniques.

First, we measure the standard deviation of the output measurements from repeated experimental runs under different repetition modes to quantify within-device variability. The noise distributions for each device are summarized using box plots to compare repetition protocols and device-specific noise profiles. Kernel density estimates (KDEs) of measurement distributions are generated for each device to visualize systematic differences beyond random noise. These plots can be generated using the Python \texttt{matplotlib} and \texttt{seaborn} packages.

Second, to analyze heterogeneity in noise behavior, K-means clustering is applied using feature vectors constructed from each experimental run, with
\[
\mathbf{x}_i = [\mu_i, \sigma_i, v_i],
\]
where $\mu_i$ is the mean, $\sigma_i$ the standard deviation, and $v_i$ the variance of measured weights for the $i$-th run. The entire dataset can be represented as
\[
X = [\mathbf{x}_1, \mathbf{x}_2, \ldots, \mathbf{x}_{N_{init}}],
\]
where $N_{init}$ is the number of the initial experiments that are included in the clustering. This multivariate approach enables clustering based not only on noise magnitude ($\sigma_i$), but also central tendency ($\mu_i$) and spread ($v_i$), providing a more complete characterization of each printer’s measurement system. The clustering is performed using the \texttt{KMeans} module from the \texttt{scikit-learn.cluster} library. When $n$ is small, setting the number of clusters $k = n$ helps identify discrete noise regimes specific to each device. However, for larger \(n\), \(k\) should be selected based on data-driven methods such as the elbow method \cite{Thorndike_1953} or silhouette score \cite{ROUSSEEUW198753} by e.g. using the \texttt{silhouette\_score} routine from the \texttt{scikit-learn} metrics library, or informed by domain knowledge, to capture meaningful noise behavior regimes while maintaining interpretability. This clustering highlights the distribution of cluster memberships across devices, reflecting device-specific operational stability.

Third, we compute pairwise divergence and distance metrics, that capture various aspects of distributional dissimilarity. These metrics include the Kolmogorov-Smirnov (KS) statistic, Kullback-Leibler (KL) divergence, Wasserstein distance, and Bhattacharyya distance using the \texttt{scipy.stats} routines and \texttt{scipy.special} \texttt{rel\_entr} module for calculating the KL Divergence, to quantify dissimilarities among devices’ noise profiles numerically. These measures, defined as follows, facilitate objective assessment of inter-device variability.

\paragraph{Kolmogorov–Smirnov (KS) Statistic}  
The KS statistic quantifies the maximum absolute difference between the empirical cumulative distribution functions (ECDFs) of two samples.  
For samples of sizes \(l\) and \(m\) with ECDFs \(F_l(x)\) and \(G_m(x)\), it is defined as  
\[
D_{\mathrm{KS}}(F_l, G_m) = \sup_x |F_l(x) - G_m(x)|.
\]
It measures the greatest deviation in cumulative probability between the two empirical distributions.

\paragraph{Kullback–Leibler (KL) Divergence}  
The KL divergence measures the relative entropy between two discrete probability distributions \(p\) and \(q\):  
\[
D_{\mathrm{KL}}(p \,\|\, q) = \sum_i p(i) \log \frac{p(i)}{q(i)},
\]
computed numerically, via the \texttt{rel\_entr} function in \texttt{scipy.special}. It is asymmetric and reflects the information loss when \(q\) approximates \(p\).

\paragraph{Wasserstein (Earth Mover’s) Distance}  
The Wasserstein-1 distance between two probability distributions \(p\) and \(q\) on \(\mathbb{R}\) is given by  
\[
W_1(p, q) = \inf_{\gamma \in \Gamma(p, q)} \int_{\mathbb{R} \times \mathbb{R}} |x - y| \, d\gamma(x, y),
\]
where \(\Gamma(p, q)\) denotes the set of all couplings with marginals \(p\) and \(q\).  
Equivalently, in terms of their cumulative distribution functions \(P(x)\) and \(Q(x)\),  
\[
W_1(p, q) = \int_{-\infty}^{\infty} |P(x) - Q(x)| \, dx,
\]
representing the minimal “mass transport” cost required to morph one distribution into the other.

\paragraph{Bhattacharyya Distance}  
The Bhattacharyya distance quantifies the degree of overlap between two probability distributions \(p\) and \(q\):  
\[
D_{\mathrm{B}}(p, q) = -\ln \!\left( \sum_i \sqrt{p(i)\, q(i)} \right),
\]
with smaller values indicating higher similarity between distributions.

Collectively, these noise analysis tools reveal significant variation and meaningful separation among noise distributions across devices, even when hardware differences are nominal. These findings justify decision criteria that favor single-device optimization when divergence metrics exceed predefined thresholds. Conversely, when the metrics indicate substantial overlap among devices, a multi-device optimization approach, pooling data across systems, becomes appropriate to improve statistical efficiency.

\subsection{Optimization Decision Criterion}
If noise distributions show little overlap, clustering analysis identifies very distinct device groups, and most pairwise divergence metrics are large, then single-device optimization is recommended: Each device should be optimized individually to account for its unique statistical behavior. Explicitly, these criteria are as follows. The KS Statistic greater than 0.5 indicates high divergence, while less than 0.2 is low. The KL Divergence greater than 5 signifies high dissimilarity; less than 2 indicates similarity. The Wasserstein Distance greater than 0.4 suggests substantial differences in distribution shape or location; less than 0.2 is low. Lastly, more negative Bhattacharyya Distance values (less than -2) mean less overlap; less negative or positive values indicate strong overlap. In contrast, when metrics are consistently low and clustering or distribution plots show devices overlap strongly, a multi-device optimization strategy is appropriate: pooling data across all devices allows efficient, unified modeling.

\subsection{Single-Device Optimization and Multi-Device Optimization}
For the Bayesian optimization (BO) implementation, we employ the Python-based software tool \texttt{gpCAM} (version 8.1.6) \cite{noack_autonomous_2020}, designed for autonomous data acquisition and analysis in materials science. Bayesian optimization (BO) is an iterative strategy for sample-efficient global optimization of black-box functions. At each iteration, a surrogate model, typically a Gaussian process (GP), is updated based on available data to approximate the underlying objective. From this model, an acquisition function is constructed, typically to balance exploration of uncertain regions and exploitation of areas promising improvement. The next experiment or evaluation point is selected by maximizing the acquisition function, guiding subsequent data collection and model refinement.

The materials science domain is characterized by large parameter spaces, heterogeneous measurement noise, and applications spanning different length scales, making surrogate modeling crucial. \texttt{gpCAM} uses GPs as surrogate models whose performance strongly depends on the kernel function and its hyperparameters (hps). 

We adopt the default anisotropic Matérn ($\nu=3/2$) kernel (once differentiable) $k_{mat}:\mathbb{R}^d\times\mathbb{R}^d\to\mathbb{R}$, which introduces $d+1$ classical hps (signal variance and $d$ length scales for a $d$-dimensional design space), in addition to noise hps, one in the single-task case and $n$ in the multi-task case. As the multi-output problem is transferred to a single-output problem with augmentation in input space, $d$ corresponds directly to the dimension of the input space for single-output optimization and to the dimension of the augmented input space for multi-output optimization. It is defined as follows
\begin{align}
	k_{mat}(x, x^{\prime}) &= \beta\left(1+\sqrt{3}\sqrt{\sum_{i=1}^d\frac{(x_i-x_i\prime)^2}{\lambda_i^2}}\right) \; \exp\left(-\sqrt{3}\sqrt{\sum_{i=1}^d\frac{(x_i-x_i\prime)^2}{\lambda_i^2}}\right),
	\label{eq:kernel}   
\end{align}
with length scale parameters $\lambda_i$ and signal variance $\beta$.
The hps are constrained with lower bounds of 0.00001, 1., 0.001 and upper bounds of 100000, 10000, 10 for signal variance, lengthscale for $f$, lengthscale for $LH$ and the noise hps are all constrained with lb=0.0001 and ub=1, with initial guesses for the first iteration set to 100, 1, 1 and 0.01 for the printer-noise related hps. For the single-task case, if the estimated lengthscales are more than one order of magnitude smaller or larger than their respective lower or upper bounds, we relax the bounds by one order of magnitude and retrain. In the following iterations, the initial guess for the hps is set to the one estimated in the previous iteration.

To select the next evaluation point, we optimize the acquisition function. For the first 11 iterations, we employed the Expected Improvement (EI) criterion $a^{EI}:\mathbb{R}^{d}\to\mathbb{R}$, defined as 
$$a^{EI}_i(x) = \mathbb{E}[I(x)]=\int_{-\infty}^{\infty}I(x)\phi(z)\;d z,$$ where $I(x)$ denotes the improvement, $\phi(z)$ is the probability density function of the normal distribution $\mathcal{N}(0,1)$, targeting improvement over the current best solution depending on point $x$ at iteration $i$. After 11 iterations, we switched to pure exploitation, maximizing the posterior mean. This pure exploitation acquisition function $a^{\mu}:\mathbb{R}^{d}\to\mathbb{R}$ can be expressed as
$$a^{\mu}_{i}(x) = \mu(x),$$ where $\mu(x)$ is the posterior mean of the surrogate model at point $x$ and iteration $i$.
A genetic algorithm with population size 50 is used for global optimization of the acquisition function, with all remaining parameters left at default values. The class \texttt{GPOptimizer} or \texttt{fvGPOptimizer} is used for training and acquisition with a loop for single- or multi-task BO, respectively.

In the single-task setting, a single point is proposed at each iteration, followed by an update of the GP before selecting the next acquisition maximum. In contrast, the multi-task setting proposes points in batches updating with the predicted values, with the batch size equal to the number of devices. Within each batch, assignments are prioritized based on the highest acquisition function value for each printer, determining the order in which points are executed for each printer.

\section{Results}\label{sec:results}
Building on our previous research in which we demonstrated a fully automated workflow for 3D printing and mechanical testing of thermoplastic specimens using parallel pellet-based printers and collaborative robotics \cite{hernandez-del-valle_robotically_2023}, we now focus on the challenges introduced by device-specific noise in such automated printer farms. In earlier work, we found significant noise variability—up to 3\% within the same printer and up to 11\% between printers when processing low molecular weight polylactide (LPLA), specifically NatureWorks Ingeo 3251D (melt flow index, MFI, 80 g/10 min) \cite{hernandez-del-valle_robotically_2023}. These levels of variability led to limited print fidelity and increased optimization costs. This finding motivates the present study, in which we investigate a network of three nominally identical Direct3D pellet printers, systematically examining the impact of noise heterogeneity and operational variability between devices. By applying our noise-aware decision-making algorithm, we are able to characterize the distinct noise profiles of each printer and demonstrate how noise-informed selection between single-device and multi-device optimization strategies mitigates these challenges. Although this case study focuses on 3D printing devices, the methodology is broadly applicable to any collection of nominally identical experimental measurement systems. Following our previous approach, the expected weight of each part is calculated, and the optimization objective is defined to minimize deviations between measured and expected weights based on the volume of the corresponding Computer-Aided Design (CAD) and the material's density:
\begin{equation}
	\Delta W = - \left| 1 - \frac{Measured\ Weight}{Expected\ Weight}\right|
\end{equation}
In the current study, we restrict optimization to two key parameters, layer height [mm] and extrusion flow multiplier [\%], as these were identified as the most influential in our earlier work.
\subsection{Initial Experimental Data Collection}
For the initial collection of experimental data, we designed a systematic sampling procedure within the parameter space allowed for each printer. Specifically, we selected the midpoint for each input parameter as a reference point:
\[
f=3000 \quad (f_{\text{lb}}=1000,\ f_{\text{ub}}=5000), \qquad \mathrm{LH}=0.4 \quad (\mathrm{LH}_{\text{lb}}=0.2,\ \mathrm{LH}_{\text{ub}}=0.6),
\]
where $f$ denotes flow rate and $\mathrm{LH}$ denotes layer height. Around this central point, we augmented the dataset by sampling an additional 19 combinations of random input parameters within these bounds, giving a total of 25 unique parameter sets. Each set was printed on all three Direct3D pellet printers, with three simultaneous repetitions per printer.

For each printed specimen, we measured the weight (in grams) as the primary quantitative outcome. This robust dataset was therefore designed to capture both process variability and device-to-device differences at the outset of our study.

\subsection{Noise Analysis Results: Distribution Analysis, Clustering and Pairwise Divergence Metrics}\label{sec:noiseana}
\begin{figure}[htbp]
	\centering
	\begin{subfigure}[b]{0.48\linewidth}
		\centering
		\includegraphics[width=\linewidth]{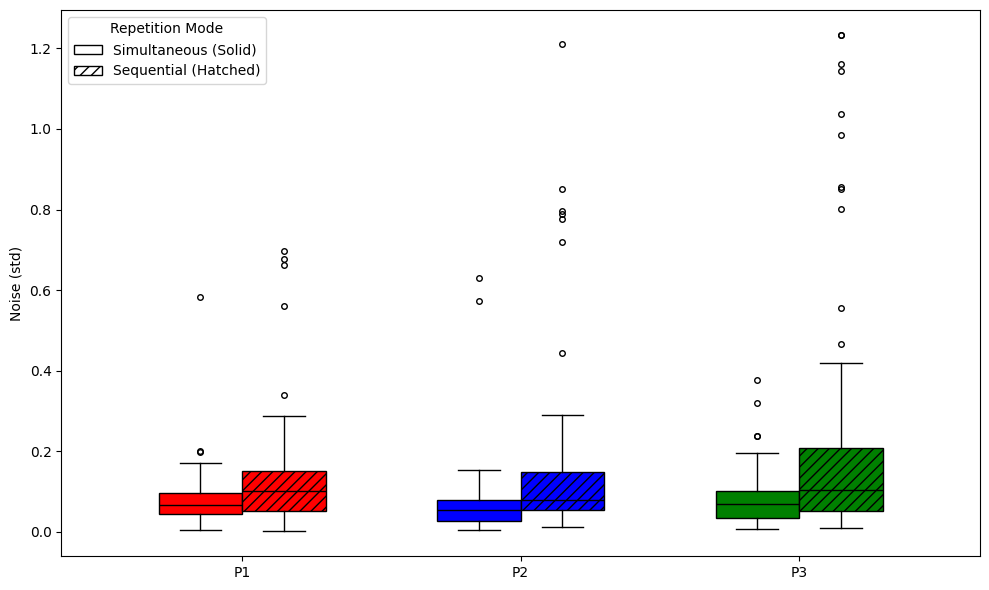}
		\caption{Box plot of noise for each printer in sequential and simultaneous repetition modes.}
		\label{fig:boxplot}
	\end{subfigure}
	\hfill
	\begin{subfigure}[b]{0.48\linewidth}
		\centering
		\includegraphics[width=\linewidth]{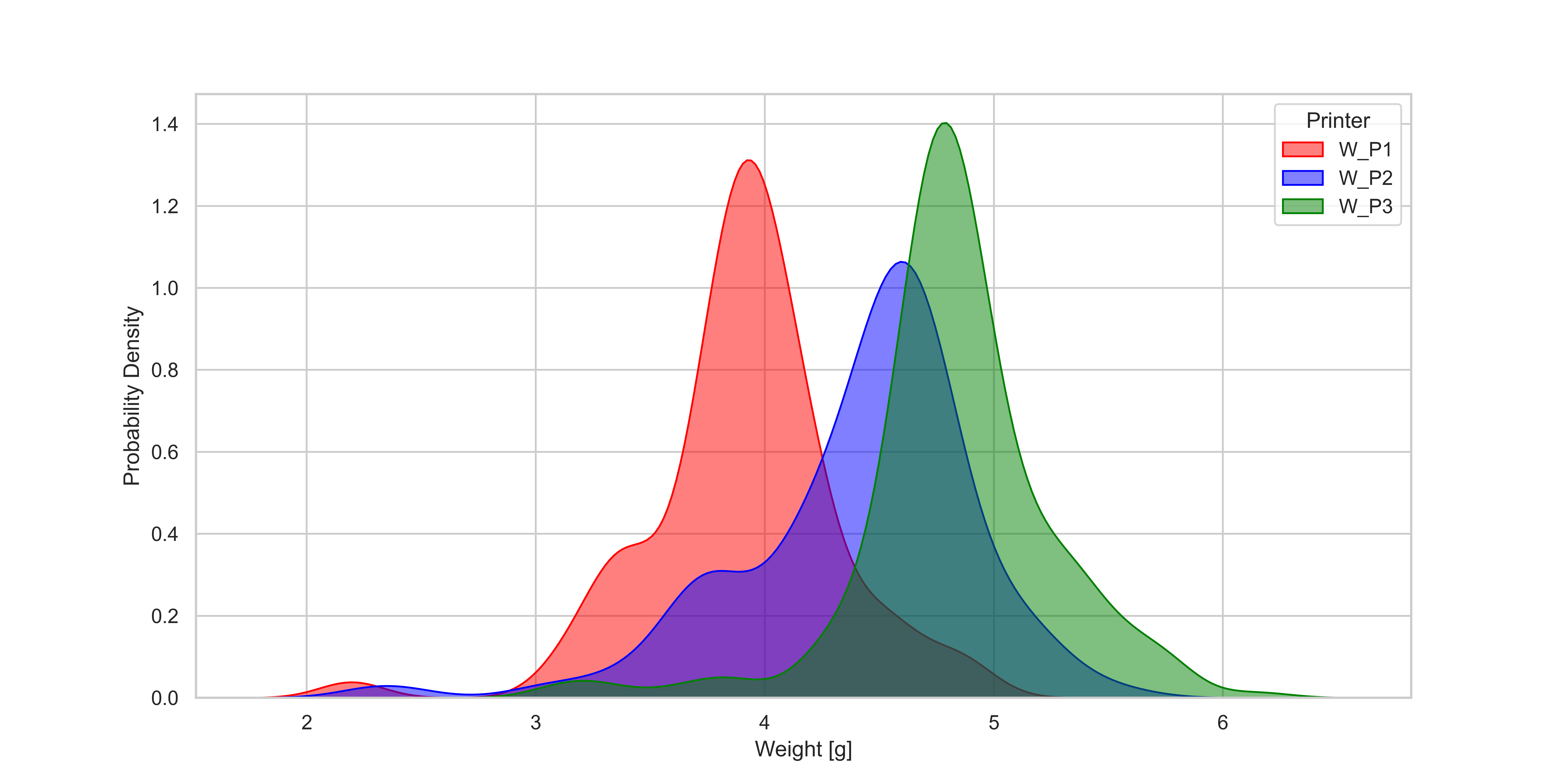}
		\caption{Kernel density estimates of print weight distributions.}
		\label{fig:kdeplot}
	\end{subfigure}
	
	\vskip\baselineskip 
	
	\begin{subfigure}[b]{0.48\linewidth}
		\centering
		\includegraphics[width=\linewidth]{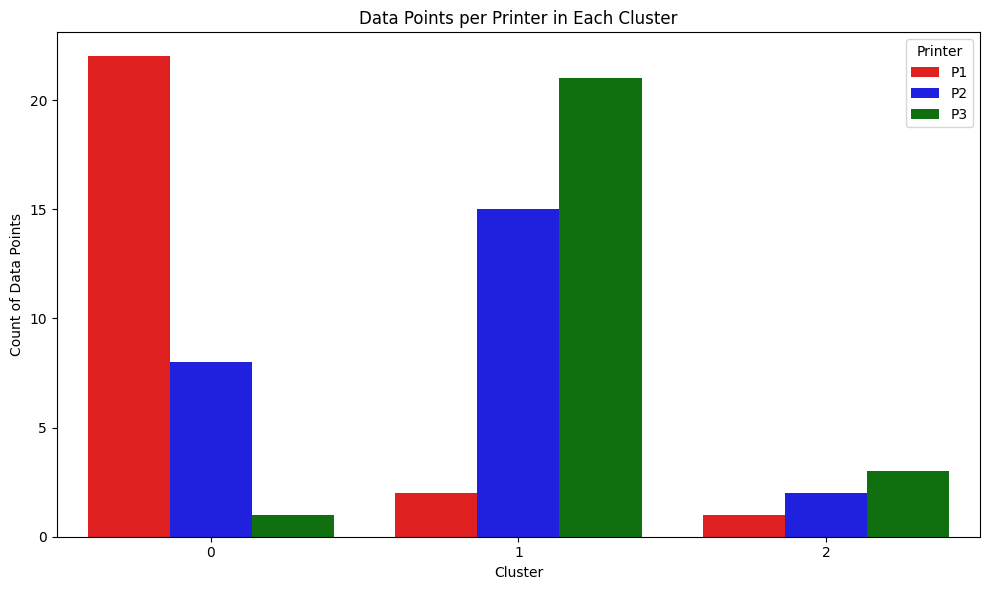}
		\caption{Cluster counts for each printer.}
		\label{fig:kmeans-counts}
	\end{subfigure}
	\hfill
	\begin{subfigure}[b]{0.48\linewidth}
		\centering
		\includegraphics[width=\linewidth]{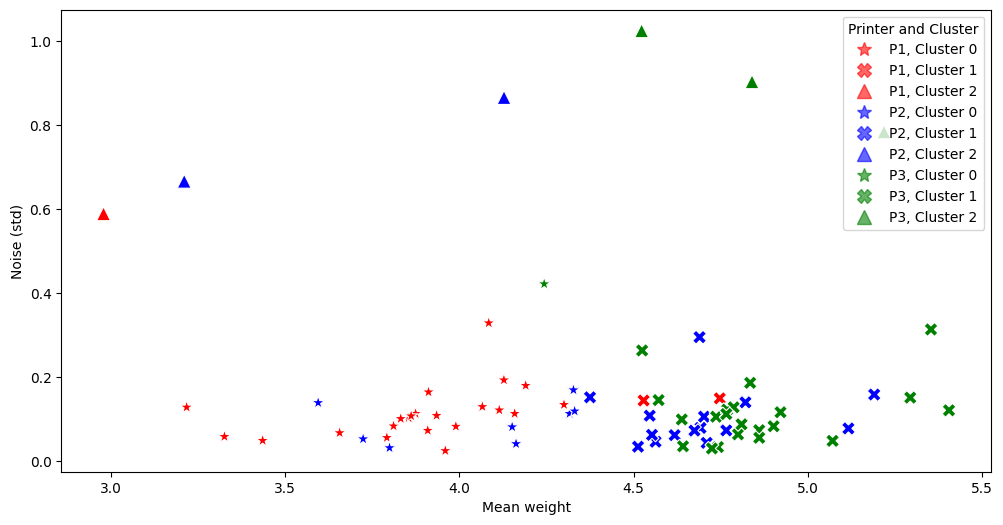}
		\caption{Noise vs. mean weight from cluster analysis.}
		\label{fig:kmeans-scatter}
	\end{subfigure}
	
	\caption{Noise and clustering analysis across printers: (a) Standard deviation of print weights per repetition mode and printer, (b) Kernel density estimates of weight distributions, (c) Cluster membership distribution by printer, (d) Noise value (standard deviation) versus mean weight for each measurement, with points colored by printer and marker shape indicating cluster assignment.}
	\label{fig:combined_noise_analysis}
\end{figure}

To characterize the variability within and between our three Direct3D pellet printers, we performed a detailed noise and cluster analysis based on the measured weights of printed specimens (cf. Figures~\ref{fig:combined_noise_analysis}). Subfigure~\ref{fig:boxplot} presents the distribution of observed noise (standard deviation of measured weights) for each printer, disaggregated by repetition mode (sequential, hatched bars; simultaneous, solid bars). Notably, each printer exhibits a distinct noise profile,with P3 demonstrating both the highest median noise and the broadest variability under both repetition protocols, while P1 consistently shows the lowest noise levels. Noise distributions from sequential repetitions generally appear broader compared to those from simultaneous runs. This is likely due to temporal variability and environmental fluctuations that accumulate over the longer duration of sequential experiments, such as gradual system drift or equipment settling times. 
Complementary distributional insights are provided in Subfigure~\ref{fig:kdeplot}, which displays kernel density estimates for the weight measurements aggregated by printer. The distributions are hardly overlapping, with W\_P1 yielding the lowest mean weights and W\_P3 the highest, despite all devices being nominally identical. Importantly, while the KDE plot visualizes differences in mean weight and overall distribution between printers, it does not reflect their respective levels of variability, which are separately quantified in the box plot. This not only confirms the presence of significant device-to-device variability but further highlights systematic performance differences that persist beyond random experimental noise.

To more deeply explore the heterogeneity of measurement noise among the three printers, we performed K-means clustering ($k=3$) on the computed mean, standard deviation and variance of measured weights for each experimental run. This feature representation allows the clustering analysis to capture subtle distinctions in both the distribution and variability of measurements.
The figure~\ref{fig:kmeans-counts} displays the count distribution of the cluster memberships for each printer. Most samples from P1 are assigned to Cluster 0, indicating consistently low noise, while the majority of P3’s samples belong to Cluster 1, reflecting its higher noise profile. P2’s data points are distributed between Clusters 0 and 1, with a notable but smaller presence in both groups. All printers have relatively few samples in the highest-noise cluster (Cluster 2).

Figure~\ref{fig:kmeans-scatter} provides a direct visualization of measurement noise separation, with each data point plotted by its mean weight and noise value, and both color and marker shape encoding printer type and cluster assignment. This representation reveals clear stratification into low (Cluster 0), intermediate (Cluster 1), and high (Cluster 2) noise regimes. High-noise data points ($\sigma > 0.6$) are almost exclusively observed in Cluster 2, predominantly arising from P2 and P3, reinforcing previous findings of greater variability in these printers. These results indicate that, despite nominal hardware identity, printers show persistent and substantial differences in their measurement noise distributions across experimental runs and clusters.
\begingroup
\small  
\setlength{\LTleft}{0pt}
\setlength{\LTright}{0pt}
\begin{longtable}{l r r r}
	\toprule 
	\textbf{Metric} & \textbf{W\_P1 vs W\_P2} & \textbf{W\_P1 vs W\_P3} & \textbf{W\_P2 vs W\_P3} \\
	\midrule 
	\endfirsthead
	
	\toprule
	\textbf{Metric} & \textbf{W\_P1 vs W\_P2} & \textbf{W\_P1 vs W\_P3} & \textbf{W\_P2 vs W\_P3} \\
	\midrule
	\endhead
	
	\bottomrule 
	\multicolumn{4}{r}{\textit{Continued on next page}} \\
	\endfoot
	
	\endlastfoot
	
	KS statistic          & 0.5867  & 0.8311  & 0.4711  \\
	Bhattacharyya distance & -2.7003 & -2.6519 & -2.5194 \\
	Wasserstein Distance  & 0.5052  & 0.9268  & 0.4224  \\
	KL divergence         & 7.9755  & 15.1497 & 5.1161  \\
	\midrule
	\caption{Comparison metrics between different pairs of W\_P samples.} \label{tab:metrics}
\end{longtable}
\endgroup

The choice between single-device and multi-device optimization depends critically on the outcomes of clustering and distribution analyses, as well as pairwise divergence metrics, specifically, KS statistics, Wasserstein distance, KL divergence, and Bhattacharyya distance. In this study, the kernel density distributions (Figure~\ref{fig:kdeplot}) reveal clear differences among devices, and clustering analyses further support distinct device-specific groupings. The pairwise divergence metrics presented in Table~\ref{tab:metrics} show high KS, Wasserstein, and KL values, together with strongly negative Bhattacharyya distances, all indicative of substantial inter-device variability. In such cases, single-device optimization is preferable, allowing each machine’s unique output distribution to be accommodated. Conversely, if divergence metrics are consistently low (values near zero and Bhattacharyya distances less negative) and the density or clustering plots exhibit substantial overlap, it is justified to employ a multi-device optimization strategy, pooling data across devices to enhance statistical efficiency.

\subsection{Single-Device Optimization vs. Multi-Device Optimization}
To illustrate the choice between single-device and multi-device optimization in this context, we present results for both approaches, even though the distributional analyses and divergence metrics clearly favor single-device optimization cf. Section~\ref{sec:noiseana}. The pronounced separation observed in the kernel density estimates (Figure~\ref{fig:kdeplot}), together with substantial pairwise divergence values, indicates that each printer operates within a distinct output regime. As such, it is most appropriate to optimize each device individually, ensuring that device-specific characteristics are properly addressed rather than pooled.
\begin{figure}[htbp]
	\centering
	\begin{subfigure}[c]{0.49\linewidth}
		\centering
		\includegraphics[width=\linewidth]{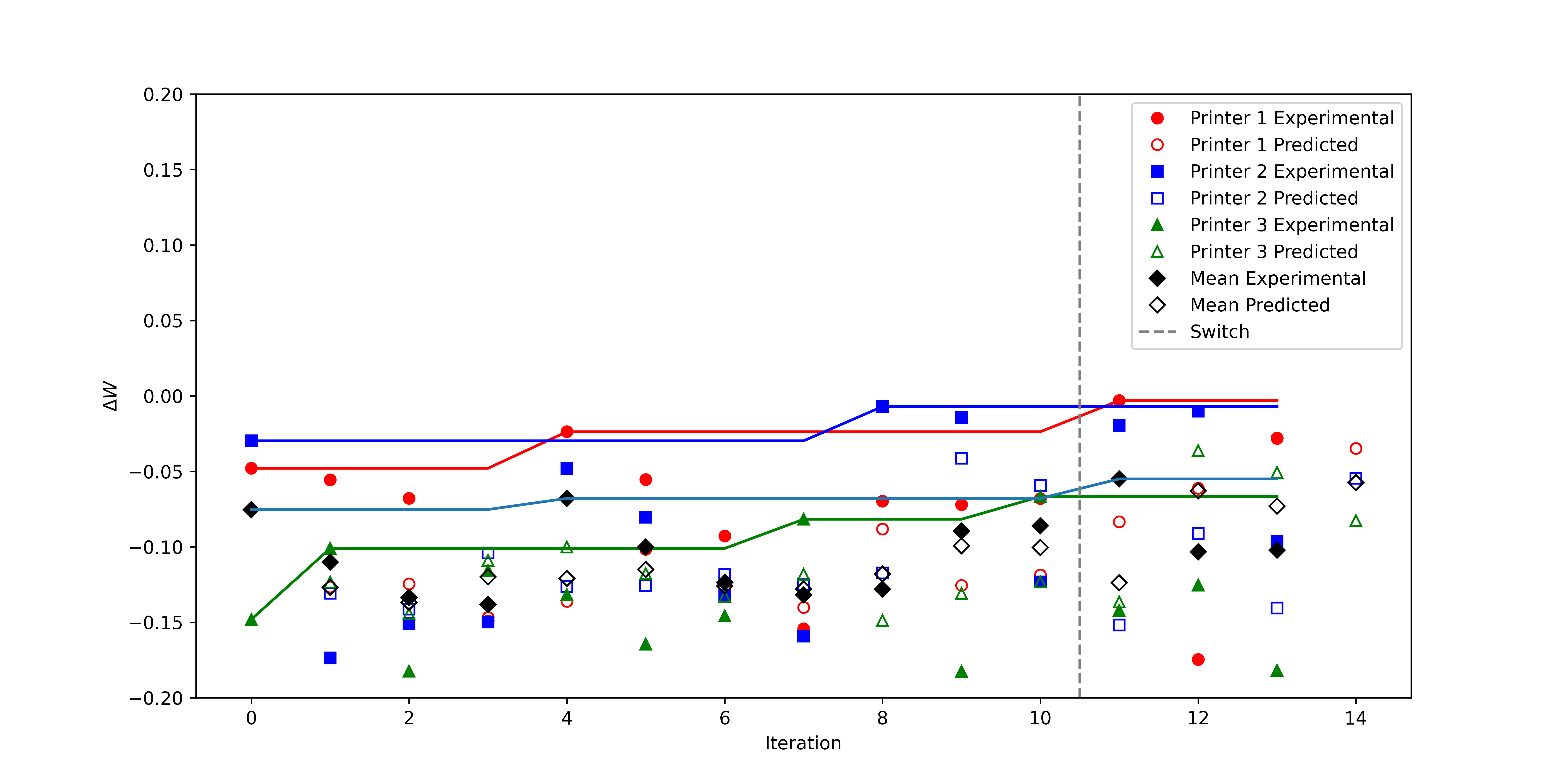}
		\caption{Convergence analysis.}
		\label{fig:multi_convergence}
	\end{subfigure}
	\begin{subfigure}[c]{0.49\linewidth}
		\centering
		\includegraphics[width=\linewidth]{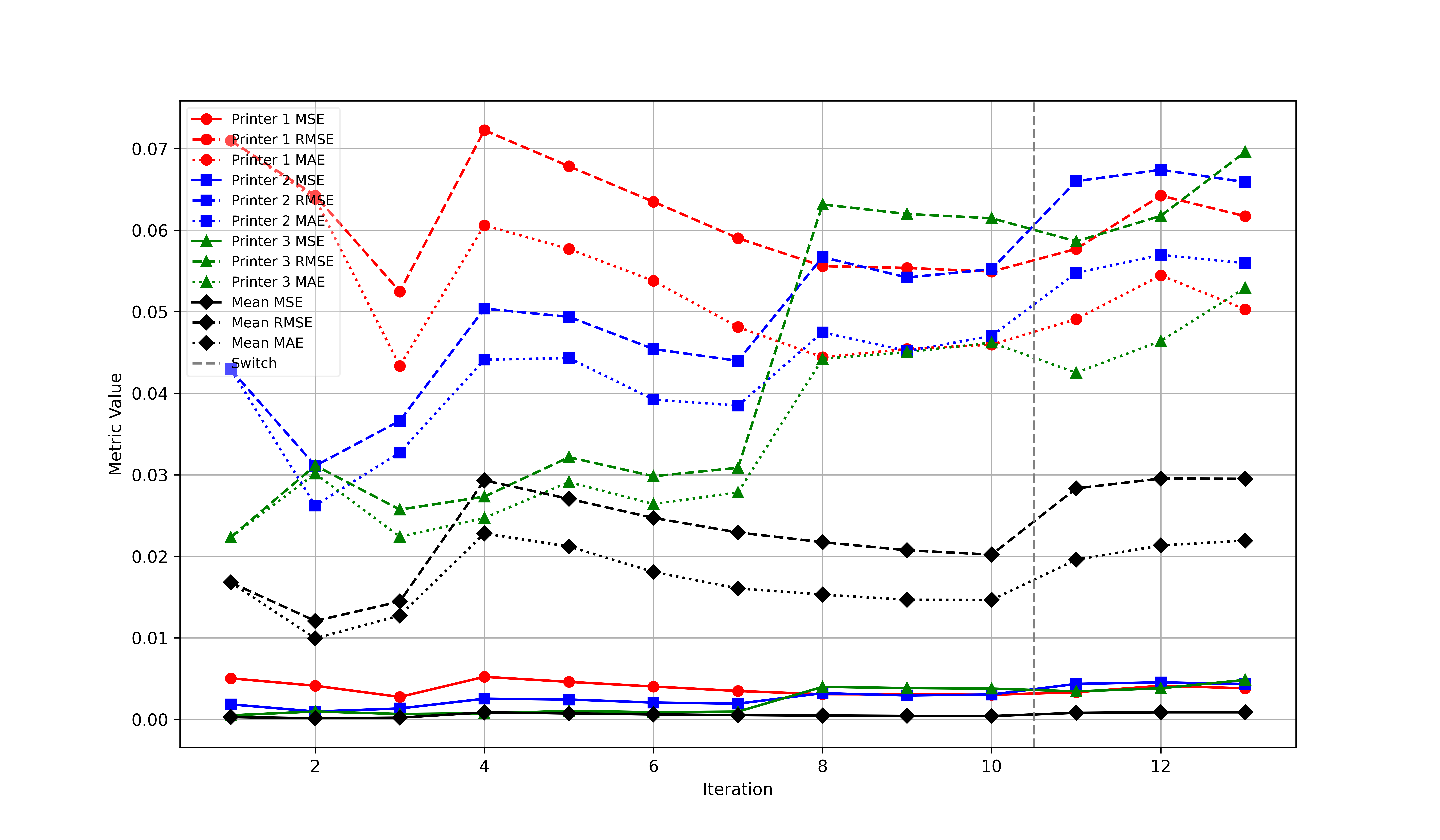}
		\caption{Performance metrics analysis.}
		\label{fig:metrics_multi}
	\end{subfigure}
	\caption{Convergence and performance analysis of multi-device optimization illustrating device-to-device variability. Robust mean performance (black) and individual performances (Printer 1: red, Printer 2: blue, Printer 3: green).}
	\label{fig:combined_convergence}
\end{figure}

To further illustrate the impact of the selected optimization strategy, Figures~\ref{fig:multi_convergence} and~\ref{fig:metrics_multi} present convergence and performance metrics behaviors for multi-device optimization. For Figure~\ref{fig:multi_convergence} we plot the evolution of weight deviation ($\Delta W$) and for Figure~\ref{fig:metrics_multi} the performance metrics with the corresponding metric value across successive iterations. In Subfigure~\ref{fig:multi_convergence}, each curve offers a distinct perspective on performance: the black curve shows mean weight deviation ($\Delta W$) averaged across all printers, the red for Printer~1, the blue for Printer~2, and the green for Printer~3.
In the multi-device optimization scheme (Figure~\ref{fig:multi_convergence}), deviations from the target weight persist for individual printers, with experimental results and model predictions exhibiting pronounced scatter and occasional systematic bias. This pooled approach, which aggregates data across devices, fails to consistently align each printer's output towards the desired target, reflecting the underlying device-to-device differences revealed in prior distributional analyses.

Subfigure~\ref{fig:metrics_multi} summarizes the convergence of error metrics (Mean Squared Error: MSE, Root Mean Squared Error: RMSE, Mean Absolute Error: MAE) during multi-device optimization for the mean and each printer using the multi-optimization approach. The joint optimization for the mean output initially decreases RMSE and MAE, but both metrics subsequently plateau or rise, particularly after switching to pure exploitation, as indicated by the dashed black line, which shows reduced predictive accuracy over time. The curves for Printer~1, Printer~2, and Printer~3 reveal similar trends, with persistent or increasing error across iterations. These results highlight that multi-device optimization does not robustly decrease prediction errors for individual devices, further reinforcing the need for device-specific approaches when systematic inter-device variability is present.
In our experimental framework, each optimization iteration queries all three printers simultaneously, producing a batch of results that is used to update the model. The device-specific bias and variability observed in the kernel density estimates (Figure~\ref{fig:kdeplot}) imply that, within each batch, predictions from printers whose outputs are closer to the target weight have a disproportionate influence on the aggregate update. This constraint, inherent to our batch evaluation protocol, creates a risk that printers with systematic offsets away from the target are undercorrected, as their less accurate outputs are masked by those with inherently lower bias. Looking forward, future multi-device optimization strategies could be improved by explicitly weighting batch updates according to device-specific performance or by applying adaptive correction factors, thereby ensuring that all printers are robustly and equitably calibrated throughout iterative learning.

\begin{figure}[htbp]
	\centering
	\begin{subfigure}[c]{0.49\linewidth}
		\centering
		\includegraphics[width=\linewidth]{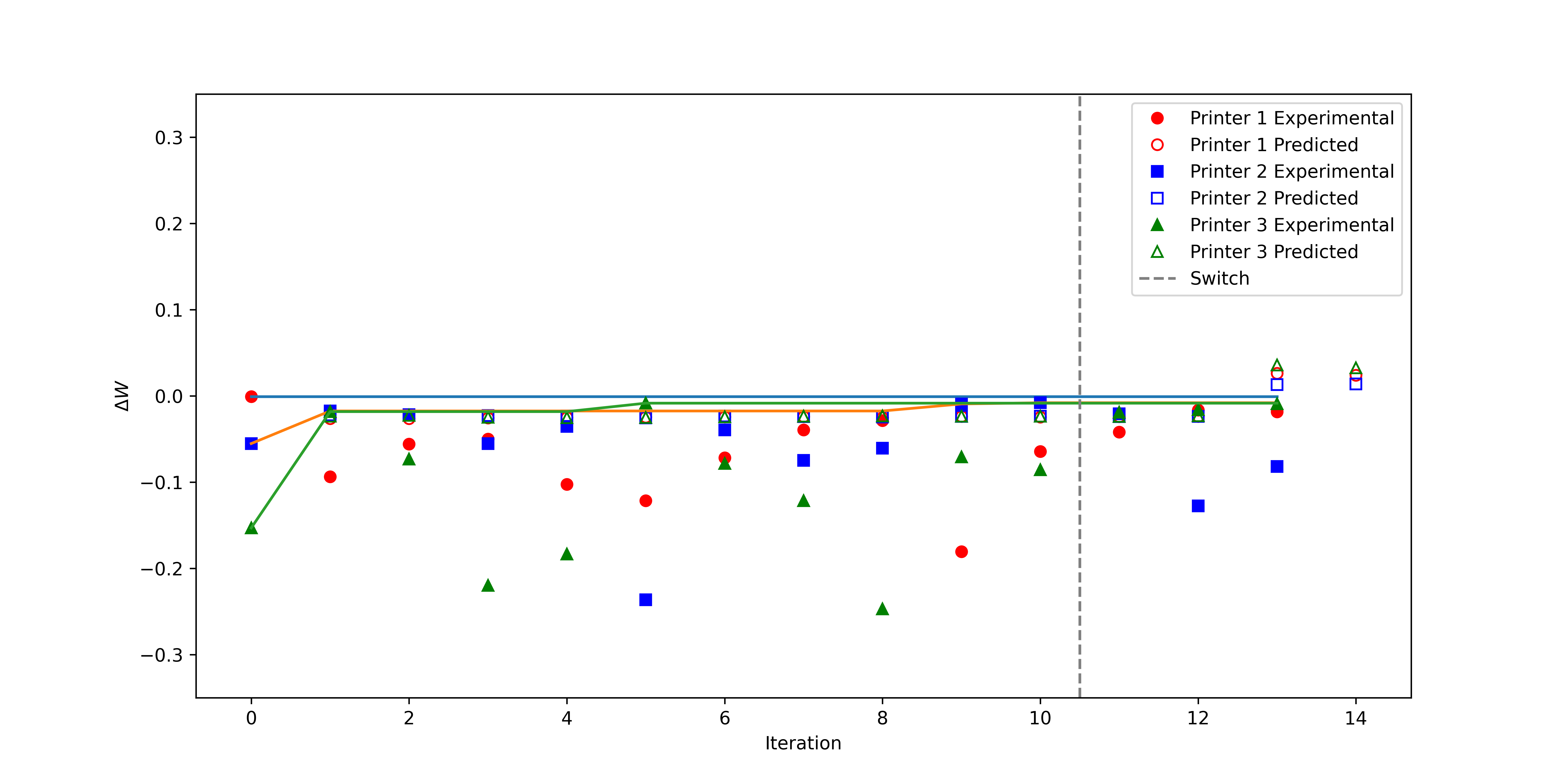}
		\caption{Convergence analysis.}
		\label{fig:single_convergence}
	\end{subfigure}
	\begin{subfigure}[c]{0.49\linewidth}
		\centering
		\includegraphics[width=\linewidth]{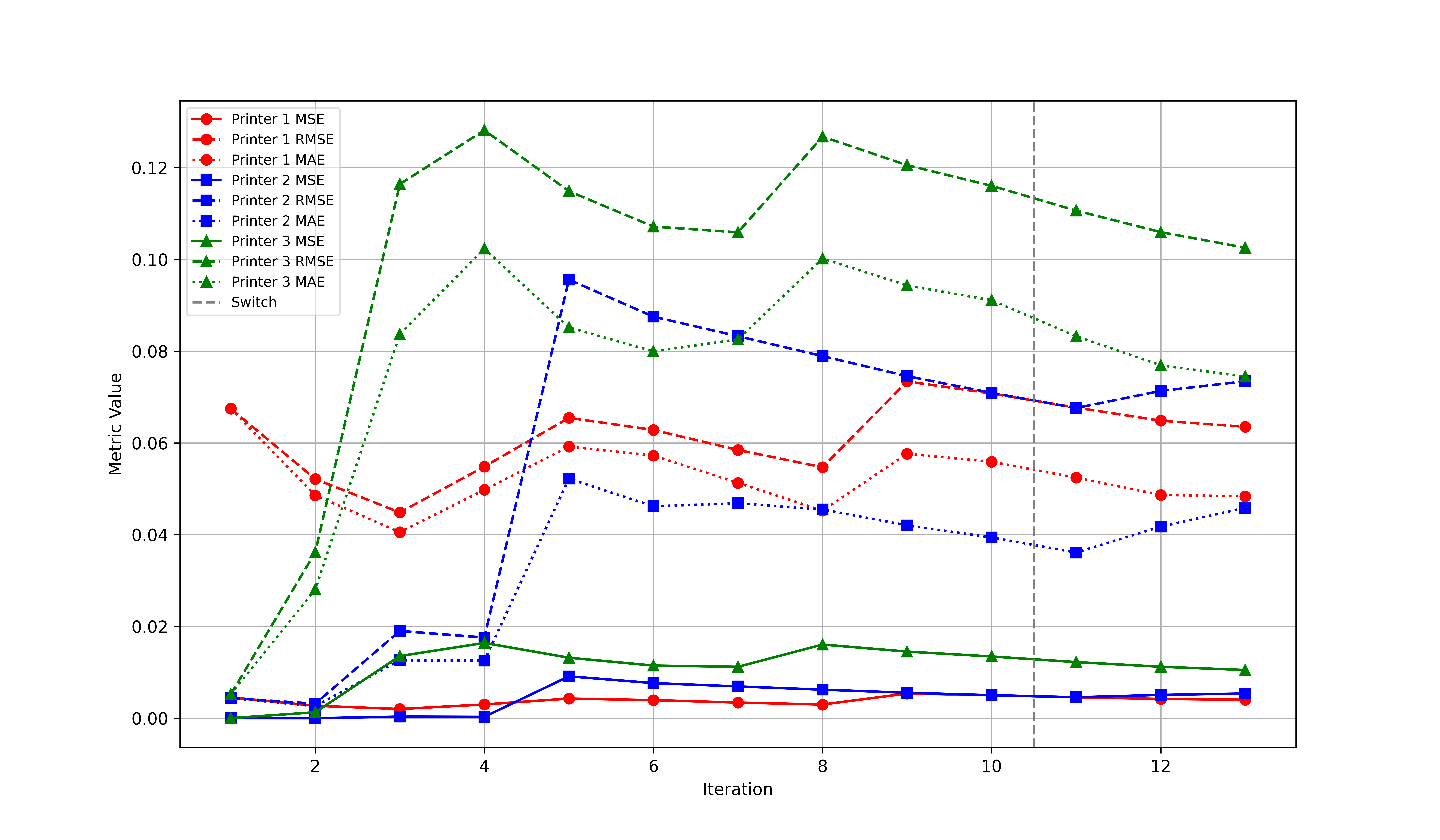}
		\caption{Performance metrics analysis.}
		\label{fig:singleopt_metrics}
	\end{subfigure}
	\caption{Convergence and performance analysis of single-device optimization illustrating device-to-device variability.}
	\label{fig:combined_metrics}
\end{figure}

In contrast, the single-device optimization results (Figures~\ref{fig:single_convergence} and ~\ref{fig:single_convergence}) demonstrate markedly improved convergence for each printer. Specifically, in Figure~\ref{fig:single_convergence}, using the same color scheme as above, the red curve highlights the optimization trajectory for Printer~1, the blue for Printer~2, and the green for Printer~3. In all cases, both experimental (filled icons) and predicted (non-filled icons) $\Delta W$ values rapidly approach zero over successive iterations, and the scatter around the target is substantially reduced compared to the multi-device optimization scheme.

Notably, for each printer, the alignment between experimental measurements and model predictions is tighter for most iterations, and systematic bias is mitigated, especially after the acquisition function switch point. These outcomes affirm that single-device optimization fully accommodates each printer's unique systematic characteristics, thereby delivering more robust and unbiased convergence. Direct comparison of individual optimization trajectories in Figure~\ref{fig:single_convergence} provides clear evidence that, in scenarios marked by significant device-specific variability, treating each printer independently is essential to achieve consistent, reproducible, and high-fidelity optimization results.

The metric convergence plots for single-device optimization (Figure~\ref{fig:singleopt_metrics}) demonstrate that handling each printer independently enables substantial and consistent reduction of error across iterations. For Printer~1, both RMSE and MAE values start at relatively high levels, reflecting notable initial deviations from the target output. As optimization progresses, these error metrics steadily decrease, apart from a slight increase for Printer~2 when switching to pure exploitation, signaling effective calibration of device-specific biases and variability. By contrast, the MSE curve for Printer~1 remains low and stable throughout, indicating that the mean squared error is consistently minimized and that the primary improvements are realized by correcting large initial deviations rather than widespread high-variance errors. In large contrast to the multi-device results, these individual curves confirm that single-device optimization successfully addresses systematic deviations and ensures that each printer approaches the target output with minimal residual error. This outcome highlights the methodological strength of treating devices independently in contexts with clear inter-device performance differences, as revealed in the underlying distributions.

In summary, these results validate the decision made under the defined criteria, analyzing noise distributions alongside cluster and pairwise divergence metrics, in this case, supporting the choice of single-device optimization over multi-device strategies under conditions of high noise variability across devices.

\section{Discussion}\label{sec:discuss}
Table~\ref{Tab:comparison} highlights the strategic trade-offs between single-device and robust multi-device optimization in high-throughput, automated measurement systems and parallel workflows such as additive manufacturing farms. While individualized optimization offers maximal precision and performance for each device, its drawbacks become acute with increasing scale, raising challenges in terms of scalability, reproducibility, maintenance, and resource efficiency. Seemingly small variability at laboratory scale may escalate dramatically at larger scales, for instance, in architectural-scale 3D printing, where noise can have tremendous structural and economic consequences \cite{WIRTH2024104081}. Conversely, robust multi-device optimization promotes platform-wide consistency and operational efficiency by generalizing across variability, but may forfeit the last increment of performance achievable through per-device customization.

The decision-making algorithm introduced in this study is expressly designed to navigate this trade-off in a principled, data-driven manner. By analyzing the distribution of noise and variability across devices, the algorithm adaptively chooses between adopting a single-device exploitation mode or a multi-device robust mode. When noise profiles are highly heterogeneous, device-specific optimization is favored to exploit unique performance characteristics. Conversely, if noise profiles are sufficiently similar, the algorithm opts for robust optimization, ensuring reproducibility, reliability, scalable management across the device fleet, and the reduction of redundant experimental effort and resource waste. This flexible and noise-aware strategy not only unifies the respective strengths of both approaches, as summarized in Table~\ref{Tab:comparison}, but also facilitates improved decision-making without sacrificing accuracy, even as conditions evolve throughout the instrument lifecycle.

A notable limitation arises as the number of devices increases substantially. The computational and analytical complexity of evaluating pairwise divergences between all device pairs grows quadratically, potentially becoming intractable for large fleets. Thus, for large-scale systems, we recommend focusing on subsets of device pairs that exhibit meaningful variability to maintain computational feasibility. Additionally, it is important to recognize that noise profiles may not remain stable over time, necessitating periodic reassessment and adaptive update mechanisms. We therefore caution that unmodeled cross-instrument noise can silently bias Bayesian optimization and erode reproducibility, and should be treated as a first-class design constraint—especially in emerging self-driving labs that rely on parallelized experimentation.

Looking to the future, the question of how best to leverage the rich information contained in device-specific noise profiles presents an intriguing direction for research and practical impact. Rather than treating noise solely as a nuisance, future strategies could use real-time noise analytics to dynamically calibrate device operation or trigger predictive maintenance routines. In particular, in high-throughput automation settings, such as industrial 3D printer farms, ongoing noise profiling could inform decisions about individualized recalibration schedules, predictive wear detection, and the optimal allocation of tasks to machines that momentarily exhibit favorable noise characteristics. By minimizing wasteful experimentation and resource-intensive trial-and-error, such 'noise-aware intelligence' integrated into high-throughput experimental automation will further close the gap between precision and scalability, driving both practical reliability and new scientific insights.
\begin{table}[h!]
	\centering
	\resizebox{\linewidth}{!}{%
	\small 
	\begin{tabular}{p{2cm}p{5.9cm}p{7.4cm}}
		\toprule
		\textbf{Criteria} & \textbf{Single-Device Optimization} & \textbf{Robust Multi-Device Optimization} \\
		\toprule
		\textbf{Pros} & Device-specific precision & Generalization \& adaptability\\
		& Higher performance per device & Resilience to noise \& variations\\
		& Easier implementation & Better long-term stability\\
		\midrule
		\textbf{Cons} & Scalability issues & Potential performance trade-offs\\
		& Lack of generalizability & Higher computational complexity\\
		& Higher maintenance required & More difficult implementation\\
		\midrule
		\textbf{Best for} & Machines with very different noise & Large-scale deployment across machines\\
		& Maximizing individual performance & Ensuring reproducibility \& robustness\\
		& Small-scale implementations & Reducing maintenance \& recalibration needs\\
		\midrule
		\textbf{Challenges} & Requires frequent recalibration & No individual best performance guarantee\\
		& May not scale efficiently & Advanced techniques required\\
		\bottomrule
	\end{tabular}}
	\caption{Comparison of Single- vs. Robust Multi-Device Optimization Approaches.}
	\label{Tab:comparison}
\end{table}

\section{Conclusions}\label{sec:concl}
The paper begins by demonstrating that device-to-device noise variability critically affects optimization in automated parallel experimental systems. It then presents a noise-aware decision-making algorithm that selects between single-device and multi-device Bayesian optimization strategies by analyzing device-specific noise profiles. Experimental validation using three nominally identical 3D printers highlights improved efficiency and reliability through these tailored optimization strategies. This approach effectively balances precision and scalability, significantly enhancing performance and reproducibility while reducing resource waste.

The study also acknowledges limitations such as the computational complexity involved in scaling to large device fleets and the need for ongoing noise profile reassessment. Importantly, it emphasizes that unmodeled cross-instrument noise can subtly bias Bayesian optimization processes and impair reproducibility. Therefore, such noise must be recognized as a critical design constraint, particularly for emerging self-driving laboratories that depend on parallelized experimentation.

Looking forward, leveraging real-time noise analytics for dynamic calibration and predictive maintenance could further advance precision and operational efficiency in high-throughput automated labs. 

Ultimately, this framework establishes a novel paradigm for noise-aware optimization in scalable experimental platforms, harmonizing the strengths of individualized and multi-device approaches to address evolving scientific and industrial challenges.

\section*{Declaration of Generative AI Use}
The authors acknowledge the use of the large language model Perplexity AI to assist in improving the clarity of sentence formulations in the manuscript. After using this tool, the authors reviewed and edited the content as needed and take full responsibility for the content of the published article.

\section*{Data and Code Availability} The experimental data files for this study are available in Zenodo \cite{dataset2025} and all the code for this work is accessible on GitHub at \cite{githubrepo}.

\section*{Author Contributions: CRediT}
The following contributions occurred: 
Conceptualization: Christina Schenk (C.S.), Miguel Hern\'andez-del-Valle (M.HdV.), Marcus Noack (M.N.), Maciej Haranczyk (M.H.); Methodology: C.S., M.HdV., M.N., M.H.; Formal analysis and investigation: C.S., M.HdV., Luis Calero-Lumbreras (L.CL.); Writing - original draft preparation: C.S.; Writing - review and editing: C.S., M.HdV., M.N., M.H.; Data Curation: L.CL., M.HdV., C.S.; Visualization: C.S., M.HdV., L.CL.; Software: C.S., M.HdV.; Supervision: C.S., M.H.; Funding acquisition: M.H., M.N.

\section*{Declaration of Competing Interest}
The authors confirm that they have no financial or personal relationships that could have influenced the work presented in this paper.

\section*{Acknowledgements}
This work was supported by the MAD2D-CM project on Two-Dimensional Disruptive Materials funded by the Community of Madrid, the Recovery, Transformation and Resilience Plan, Spain, and NextGenerationEU from the European Union. Marcus Noack’s contribution to this work was supported by the Center for Advanced Mathematics for Energy Research Applications (CAMERA), funded jointly by the Advanced Scientific Computing Research (ASCR) and Basic Energy Sciences (BES) programs in the U.S. Department of Energy’s Office of Science, under Contract No. DE-AC02-05CH11231.

\bibliographystyle{unsrt}
\bibliography{sample}

\end{document}